\documentstyle[aps,pra,epsfig,twocolumn,floats]{revtex}
\begin{document}
\draft
\title{
Sampling the canonical phase from phase-space functions
}
\author{J. Fiur\'{a}\v{s}ek,$^{1,2}$
M. Dakna,$^{3}$
T. Opatrn\'{y},$^{4,2}$
and D.--G. Welsch$^{4}$
 }
\address{
$^{1}$ Department of Chemical Physics, Weizmann Institute of Science,
76100 Rehovot, Israel \\
$^{2}$ Faculty of Science, Palack\'{y} University, 
Svobody 26, CZ-77146 Olomouc, Czechia \\
$^{3}$ Institut f\"{u}r Theoretische Physik, 
 Universit\"at   G\"{o}ttingen,   Bunsenstr. 9, D-37073 Germany \\
$^{4}$ Theoretisch-Physikalisches Institut,
Friedrich-Schiller-Universit\"at, Max-Wien-Platz 1, D-07743 Jena,
Germany
}
\date{\today}

\maketitle

\begin{abstract}

We discuss the possibility of sampling exponential moments of the canonical
phase from the $s$-parametrized phase space functions. We show that the
sampling kernels exist and are well-behaved  for any $s$ $>$ $-1$, whereas for 
$s$ $=$ $-1$ the kernels diverge in the origin. 
In spite of that we show that the phase space moments can be sampled with any
predefined accuracy from the $Q$-function measured in the double-homodyne
scheme with perfect detectors. We discuss the effect of imperfect
detection and address sampling schemes using other measurable
phase-space functions. Finally, we discuss the problem of sampling the 
canonical phase distribution itself.

\end{abstract}


\section{Introduction}

Studying the role of phase in quantum mechanics has a long history (for a
review on the phase concepts, see, e.g., \cite{Perinas}).  Its importance in
today's problems is also apparent. For example, phase in atomic systems has
recently been used for storing quantum information \cite{Ahn}, and phase and
photon number measurements have been considered as a basis in some quantum
teleportation  schemes \cite{Milburn}. Notwithstanding the various
phase-dependent effects in quantum physics, phase itself has not been uniquely
measured and its very definition as physical quantity has been subject to many
disputes. Whereas for highly excited (quasi-classical) states different
approaches give similar  results, the various concepts differ in the phase
properties of quantum states  close to vacuum. Therefore the question has been
arisen of  what are the differences between these approaches and how relevant 
are they experimentally.  In this paper we concentrate  on the canonical phase
and its relation to $s$-parametrized phase-space functions, with special
emphasis on  the measurability of its exponential moments by ``weighted''
averaging of measured phase space functions. 

The canonical phase distribution $P(\varphi)$ of a radiation field mode
(harmonic oscillator) prepared in a quantum state described by a density 
operator $\hat \varrho$ is defined by
\begin{eqnarray}
\label{1}
P(\varphi) = (2\pi)^{-1}\langle \varphi |\hat \varrho | \varphi \rangle,
\end{eqnarray}
where the Fock state expansion of the (unnormalizable) phase states 
$|\varphi \rangle$ reads
\begin{eqnarray}
\label{2}
|\varphi \rangle = \sum_{n=0}^{\infty} e^{in\varphi} |n\rangle.
\end{eqnarray}
Even though there has been no known experimental scheme 
that is directly governed by $P(\varphi)$, this distribution
has very nice properties: it is 
non-negative, conjugated to the photon-number
distribution (in the sense that a phase shifter shifts a phase
distribution while a number shifter does not change it \cite{LVBP95}), there
exist number-phase uncertainty relations \cite{uncert},
and in comparison to other phase distributions, $P(\varphi)$
is the most sharp one.

The lack of direct experimental availability of $P(\varphi)$ has led us to the
search of schemes for sampling the canonical phase statistics from quantities
that can be measured directly \cite{ODW98,DOW98,DBMOSW98}. In balanced homodyne
detection (for a review on quantum state measurement using homodyning, see,
e.g., \cite{WVO}), the exponential moments  $\Psi_k$ of the canonical phase,
\begin{eqnarray}
\label{defpsik}
\Psi_k = \int_{2\pi} d\varphi\, e^{ik\varphi} P(\varphi), 
\quad
\Psi_{-k} = \Psi_k^\ast,
\end{eqnarray}
can be sampled by integrating the measured 
quadrature-component statistics multiplied by well-behaved kernel 
functions \cite{ODW98,DOW98,DBMOSW98}. 
An advantage of the method is that it applies to both the quantum 
regime and the classical regime in a unified way. 
Of course, the question has been as of whether or not it is 
possible to find other (and possibly better) measurement schemes 
suitable for sampling the exponential moments of the canonical phase. 

It is well known that balanced double-homodyne detection 
(eight-port homodyning, \cite{Noh}) provides us with a two dimensional 
set of data whose statistics correspond to a $s$-parametrized 
phase-space function $W_{s}(q,p)$ with $s$ $\!\leq$ $\! -1$
\cite{Freyberger}. In this scheme,
the limiting case of \mbox{$s$ $\!=$ $\!-1$}, which corresponds to 
the Husimi $Q$-function $Q(q,p)$ $\!=$ $\!W_{-1}(q,p)$, requires
perfect detection, i.e., 100\% detection efficiency. Having a sampling 
scheme leading from a measured $s$-parametrized phase-space function to 
the exponential canonical-phase moments would be the most direct method of 
measuring the exponential moments of the canonical phase. Since each 
measurement event $(q,p)$ already yields a phase value Arg$(q+ip)$, 
the measured values only need to be ``weighted'' by the kernel functions 
in the averaging procedure yielding the exponential moments $\Psi_k$.

There are also measuring schemes, e.g., unbalanced homodyning, suitable
for determining $s$-parametrized phase-space functions $W_{s}(q,p)$ 
with larger values of $s$ 
\cite{Wallentowicz96}. However, in these schemes 
the functions $W_{s}(q,p)$ are not obtained in terms of the statistics of 
measurement events $(q,p)$, but they are obtained pointwise for 
each phase-space point $(q,p)$ set up in the experiment. Moreover, 
they are typically reconstructed from the measured data rather
than measured directly. Nevertheless, it is interesting to ask the 
question of the prospects of phase measurement in schemes of that type. 

In this paper, we try to answer the questions raised above, focusing 
our attention to the problem of using balanced double-homodyne
detection for sampling the exponential moments of the canonical 
phase. In Sec.~\ref{Sec-moments} we present the kernels
that relate the $s$-parametrized phase-space functions to the
exponential phase moments, and in Sec.~\ref{Sec-homodyne}
we apply the results to direct sampling of the exponential 
phase moments in balanced double-homodyne detection. 
Other measurement schemes are discussed in Sec.~\ref{Other}.
Section \ref{Sec-distr} addresses the problem of determining 
the phase distribution itself, and a conclusion is given
in Sec.~\ref{Sec-conclusion}.


\section{The kernel function}
\label{Sec-moments}

Our task is to find the kernel function $K_k(r;s)$ such that the
exponential moments of the canonical phase can be given by ($k$ $\!>$ $\!0$)
\begin{eqnarray}
\lefteqn{
\Psi_k = \big\langle\hat{E}^k\big\rangle
}
\nonumber\\&&\hspace{4ex}
 =\int_0^{2\pi}d\varphi\, e^{ik\varphi}
 \int_0^{\infty}r dr \, W(r,\varphi;s)K_k(r;s),
 \label{PSI}
\end{eqnarray}
and $\Psi_{-k}$ $=$ $\Psi_k^{*}$,
where
\begin{eqnarray}
\label{3}
\hat{E} = \sum_{n=0}^\infty |n\rangle\langle n+1|.
\end{eqnarray}
In Eq.~(\ref{PSI}), the phase-space function $W(r,\varphi;s)$ is written 
in polar coordinates, i.e., $W(r,\varphi;s)$ $\!=$ 
$\!W_{s}(r\cos \varphi,r \sin \varphi)$. Note that $e^{ik\varphi}K_k(r;s)$ 
is the $(-s)$-parametrized phase-space function of the operator $\hat E^k$.
We now take advantage of the expression \cite{DOW98}
\begin{equation}
  \Psi_k
 = \sum_{l=0}^{\infty}\sum_{n=0}^{l} \frac{(-1)^{l-n}}{(l\!-\!n)!
 \sqrt{n!(l\!+\!n)!}} \langle \hat a^{\dag l}\hat a^{l+k} 
 \rangle ,
 \label{PSIaa}
\end{equation}
where the expectation value of the normally ordered correlations 
of the photon creation and destruction operators can be calculated 
by means of $W(r,\varphi;s)$ as \cite{Qstat}
\begin{eqnarray}
\lefteqn{
 \langle \hat a^{\dagger l} \hat a^{l+k}\rangle=
 (-1)^l l! \left(\frac{1-s}{2}\right)^l
 \int_0^{2\pi} d\varphi\, e^{ik\varphi} 
}
 \nonumber \\&&\hspace{6ex}
 \times
 \int_0^{\infty} rdr \,  r^k L_l^k
 \left(\frac{2r^2}{1-s}\right) W(r,\varphi;s)
 \label{MOMENTSA}
\end{eqnarray}
($L_l^k$, Laguerre polynomial). Combining Eqs.~(\ref{PSI}), (\ref{PSIaa}),
and (\ref{MOMENTSA}), we derive (Appendix \ref{derivkernel})
\begin{eqnarray}
\lefteqn{
 K_k(r;s)=\frac{r^k 2^{k+1}}{\pi^{k/2}}
 \int_{0}^{\infty} d\rho \,  \Bigg\{ \rho^{k\!-\!1} \Omega^{(k)}(\rho^2)
} 
 \nonumber \\&&
 \times
 \left[1\!+\!s\!+\!(1\!-\!s)e^{-\rho^2}\right]^{-\!k\!-\!1}
 \exp\!\left[-\frac{2(1\!-\!e^{-\rho^2})r^2}
 {1\!+\!s\!+\!(1\!-\!s)e^{-\rho^2}}\right]
 \Bigg\}.
 \nonumber \\&& 
  \label{KERNEL}
\end{eqnarray}
Here, the function $\Omega^{(k)}(\rho^2)$ is given by
\begin{equation}
 \Omega^{(k)}(\rho^2)=e^{-\rho^2}\sum_{n=0}^{\infty}\frac{(-1)^n}{n!}
 A_{n}^{(k)}\rho^{2n},
 \label{OMEGATAYLOR}
 \end{equation}
where
 \begin{eqnarray}
 \label{5}
\lefteqn{ 
A_{n}^{(k)} = \int_0^\pi d\varphi_1\,\sin^{k-2}\varphi_1\ldots
}
\nonumber\\&&\ldots 
 \int_0^\pi d\varphi_i\,\sin^{k-i-1}\varphi_i
 \ldots
 \int_0^{2\pi} d\varphi_{k-1}\,\Big\{
 \left[\sin^2\varphi_1 \right.
\nonumber\\&& \hspace{6ex}\times \left.
 (1+\sin^2\varphi_2(1+\ldots(1+\sin^2\varphi_{k-1})))
 \right]^n \Big\}.
\end{eqnarray}
It is worth noting that $K_k(r;s)$ is {\em unique\/}, which follows from the 
fact that $K_k(r;s)$ is the phase-space function of $\hat{E}^k$ and from 
the uniqueness of phase-space representations. This is in contrast
to the kernel functions that relate quantities to the 
quadrature-component statistics measured in balanced homodyne scheme,
where certain functions can be added to the kernels without
changing the result \cite{DOW98,WVO,DAriano}.

\begin{figure}[!t]
\noindent
\centerline{\epsfig{figure=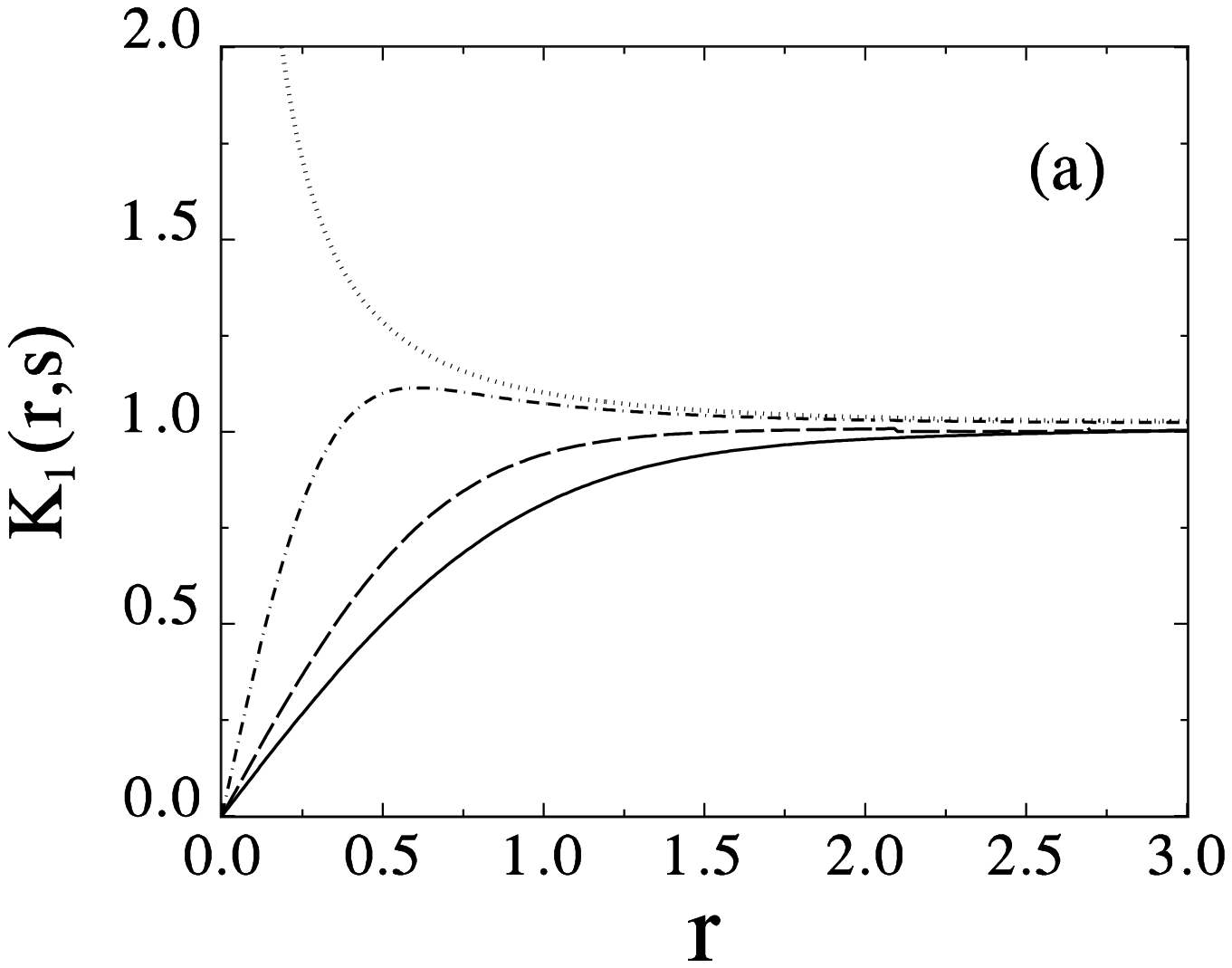,width=0.7\linewidth}}

\vspace*{2mm}

\centerline{\epsfig{figure=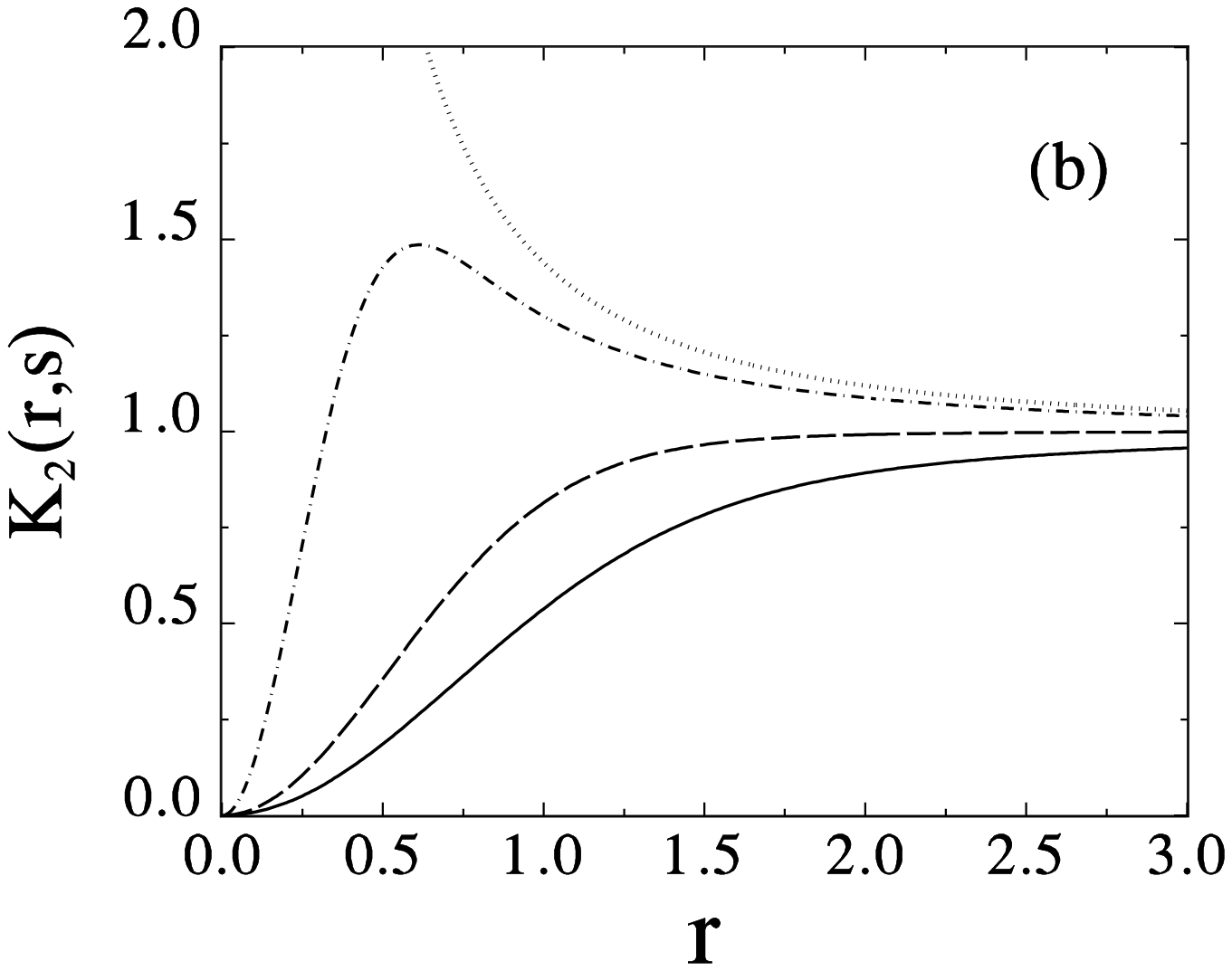,width=0.7\linewidth}}

\vspace*{2mm}

\centerline{\epsfig{figure=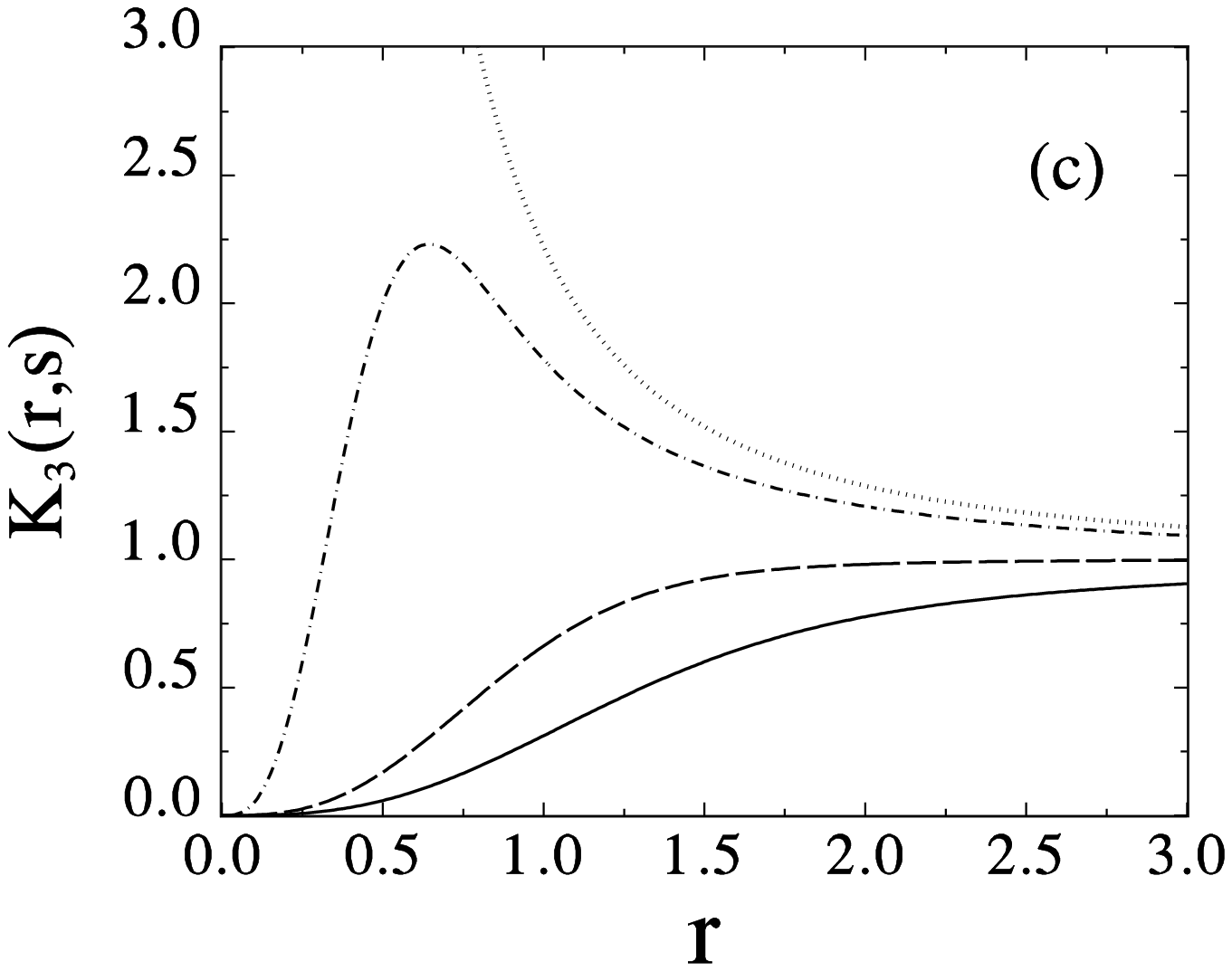,width=0.7\linewidth}}

\vspace*{2mm}

\caption{ 
The kernel function $K_k(r;s)$ for $k$ $\!=$ $\!1,2,3$ and 
\mbox{$s$ $\!=$ $\!0.75$} (full line), \mbox{$s$ $\!=$ $\!0$} (broken line),
\mbox{$s$ $\!=$ $\!-0.75$} (dash-dotted line),
\mbox{$s$ $\!=$ $\!-1$} (dotted line).
}
\label{fig1}
\end{figure}

The integral in Eq.~(\ref{KERNEL}) converges for $s$ $\!>$ $\!-1$ because
\begin{eqnarray}
|\Omega^{(k)}(\rho^2)|<e^{-\rho^2}V_k,
\end{eqnarray}
$V_k$ being some constant. 
Plots of the kernel function for different 
values of $s$ and $k$ are shown in Fig.~\ref{fig1}. We can see that 
$K_k(r;s)$ monotonically increases with $r$ from zero to one 
for $s$ $\!\ge$ $\!0$. 
If $s$ $\!<$ $\!0$, then $K_k(r;s)$ attains the maximum
at a finite value of $r$. The position of the maximum shifts towards 
the origin and the value of the maximum tends to infinity as $s$ 
$\!\rightarrow$ $\!-1$. Hence the kernels that relate
the exponential phase moments to the $Q$-function diverge at
$r$ $\!=$ $\!0$. To be more specific, it can be shown
(Appendix B) that
\begin{eqnarray}
K_k(r;-1) \propto r^{-k}
 \label{kerndiv}
\end{eqnarray}
near the origin.

Though the function $K_k(r)$ $\!\equiv$ $\!K_k(r;-1)$ diverges, 
it can be used to obtain the exponential phase moments $\Psi_k$ 
from the $Q$-function $Q(r,\varphi)$ $\!=$ $\!W(r,\varphi,-1)$. It
is not difficult to prove that Eq.~(\ref{PSI}) can be rewritten as
\begin{equation}
 \Psi_k = \int_0^\infty rdr \, Q_k(r) K_k(r) , 
 \label{PSIINT2}
\end{equation}
where
\begin{eqnarray}
\lefteqn{
 Q_k(r) =  \int_0^{2\pi} d\varphi \, e^{ik\varphi}Q(r,\varphi)
} 
 \nonumber \\&&\hspace{6.5ex} 
  =\, 2e^{-r^2}\sum_{n=0}^{\infty}\frac{r^{2n+k}}{\sqrt{n!\,
 (n+k)!}}\rho_{n+k,n}
 \label{Qrho}
\end{eqnarray}
($\rho_{n+k,n}$ $\!=$ $\!\langle n$ $\!+$ $\!k|\hat{\varrho}|n\rangle$).
It follows from Eq.~(\ref{Qrho}) that $Q_k(r)$ $\!\propto$ $\!r^k$ for
small $r$, and thus $Q_k(r)$ exactly compensates for the
divergence of $K_k(r)$, Eq.~(\ref{kerndiv}). In other words,
if the $Q$-function of the state is known 
exactly,
then the integration in (\ref{PSIINT2})
can be performed straightforwardly, thus yielding the sought $\Psi_k$.
However, measurement of the $Q$-function is always associated
with some error, so that the region close to the origin of the
phase space needs careful consideration in praxis.


\section{Canonical phase from double homodyning}
\label{Sec-homodyne}

\subsection{Statistical error}
\begin{figure}[!t!b]
\centerline{\epsfig{figure=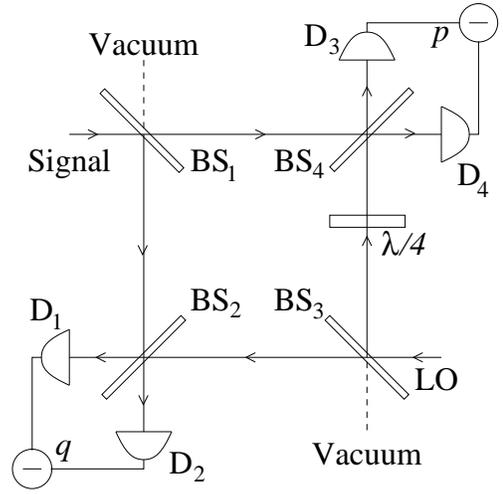,width=0.75\linewidth}}
\vspace*{3mm}
\caption{ 
Double-homodyne scheme \protect\cite{Noh}. 
The signal beam is split on a beam splitter BS$_1$ and
the resulting beams are mixed with strong coherent local oscillator (LO) on the
beam splitters BS$_2$ and BS$_4$. The LO beams at  BS$_2$ and BS$_4$ stem from a
common source, split at BS$_3$, and their phases differ by $\pi/2$, determined
by the $\lambda /4$ phase shifter. The difference of photocurrents measured
at the detectors D$_1$ and D$_2$ is proportional to $q$ and the
photocurrent difference at D$_3$ and D$_4$ is proportional to $p$.
}
\label{fig8port}
\end{figure}

Let us consider balanced double-homodyne detection 
(Fig.~\ref{fig8port}) and first assume perfect detection. 
Each experimental event then gives a pair of real numbers 
which, after rescaling, define a point in the phase space of the 
signal, and the probability density of detecting the space points is 
equal to the $Q$-function of the signal state \cite{Freyberger}.
When the $j$th measurement yields the
phase-space point with polar coordinates $(r_j,\varphi_j)$ and 
altogether $N$ measurements are performed, then the
exponential phase moments
can be estimated to be
\begin{eqnarray}
 \Psi_k^{({\rm est})} = \frac{1}{N}\sum_{j=1}^{N} \exp (k\varphi_{j})
 K_{k}(r_{j}).
 \label{PSIEST}
\end{eqnarray}

In order to answer the question of how close is $\Psi_k^{({\rm est})}$
to the
actual moment $\Psi_k$, we calculate the mean value and dispersion of the 
estimate (\ref{PSIEST}) over all possible measurement results. 
Since individual measurement outcomes are independent of each other, 
we can take advantage of the summation rule for mean values and 
dispersions of independent quantities. Thus, for the real part
of $\Psi_k^{(\rm est)}$ we get the mean value
\begin{eqnarray}
\lefteqn{
E\!\left({\rm Re}\, \Psi_k^{(\rm est)}\right)=\frac{1}{N}\sum_{j=1}^{N} 
 E\!\left[ \cos(k\varphi_{j})K_{k}(r_{j})\right] 
} 
\nonumber \\&&\hspace{2ex}
 =\,\frac{1}{N}\sum_{j=1}^{N}
 \int_{2\pi} \!\!d\varphi_{j} \int_0^{\infty}\!\!\! r_{j}dr_{j} \,
 \cos (k\varphi_{j}) K_{k}(r_{j}) Q(r_{j},\varphi_{j})
\nonumber \\&&\hspace{2ex}
 =\, \frac{1}{N}N \ {\rm Re}\, \Psi_k = {\rm Re}\, \Psi_k ,
\end{eqnarray}
as it should be, and the dispersion
\begin{eqnarray}
\lefteqn{
 D\!\left({\rm Re}\Psi_k^{({\rm est})}\right) = \frac{1}{N^2}\sum_{j=1}^N
 D\!\left[\cos (k\varphi_j) K_k(r_j)\right]
}
\nonumber \\&&\hspace{2ex}
 =\, \frac{1}{N} \Bigg\{  \int_{2\pi}d\varphi \int_{0}^{\infty} rdr \,
 \cos ^2 (k\varphi) K_k ^2 (r) Q(r,\varphi) 
\nonumber \\ &&\hspace{2ex}
 - \left[  \int_{2\pi} d\varphi \int_{0}^{\infty} rdr \,
 \cos (k\varphi)
 K_k (r) Q(r,\varphi) \right] ^2
 \Bigg\} .
 \label{DSTAT}
\end{eqnarray}
Similar expressions hold for the imaginary part of 
$\Psi_k^{(\rm est)}$. 
Since cos$^2(k\varphi)$ $=$ $1/2$ $+$ cos$(2k\varphi)/2$,
after performing the angular integration in the first term on the
right-hand side of Eq.~(\ref{DSTAT}),
the radial part contains the product $Q_0(r)K_k ^2 (r)$
so that this integral over the
divergent kernel can become infinite. 
Let $n_0$ be the number of photons
at which the Fock expansion of the state starts. Taking into account 
Eq.~(\ref{Qrho}), we see that the integrand behaves as $\propto$
$\!r^{2(n_0-k)+1}$ for small $r$. Thus, the exponential phase moments 
$\Psi_k$ can be directly sampled from the double-homodyne data, provided
that $k$ $\!<$ $\!n_0$ $\!+$ $\!1$, because in this case the dispersion 
of the estimation is bounded. In the opposite case of 
\mbox{$k$ $\!\ge$ $\!n_0$ $\!+$ $\!1$}, the statistical fluctuation 
diverges so that the exponential phase moments cannot be sampled 
without a proper regularization of the kernels. 
Note that for states that contain the vacuum, regularization of
the kernels is necessary for all exponential phase moments. 

\begin{figure}[!t!]
\centerline{\epsfig{figure=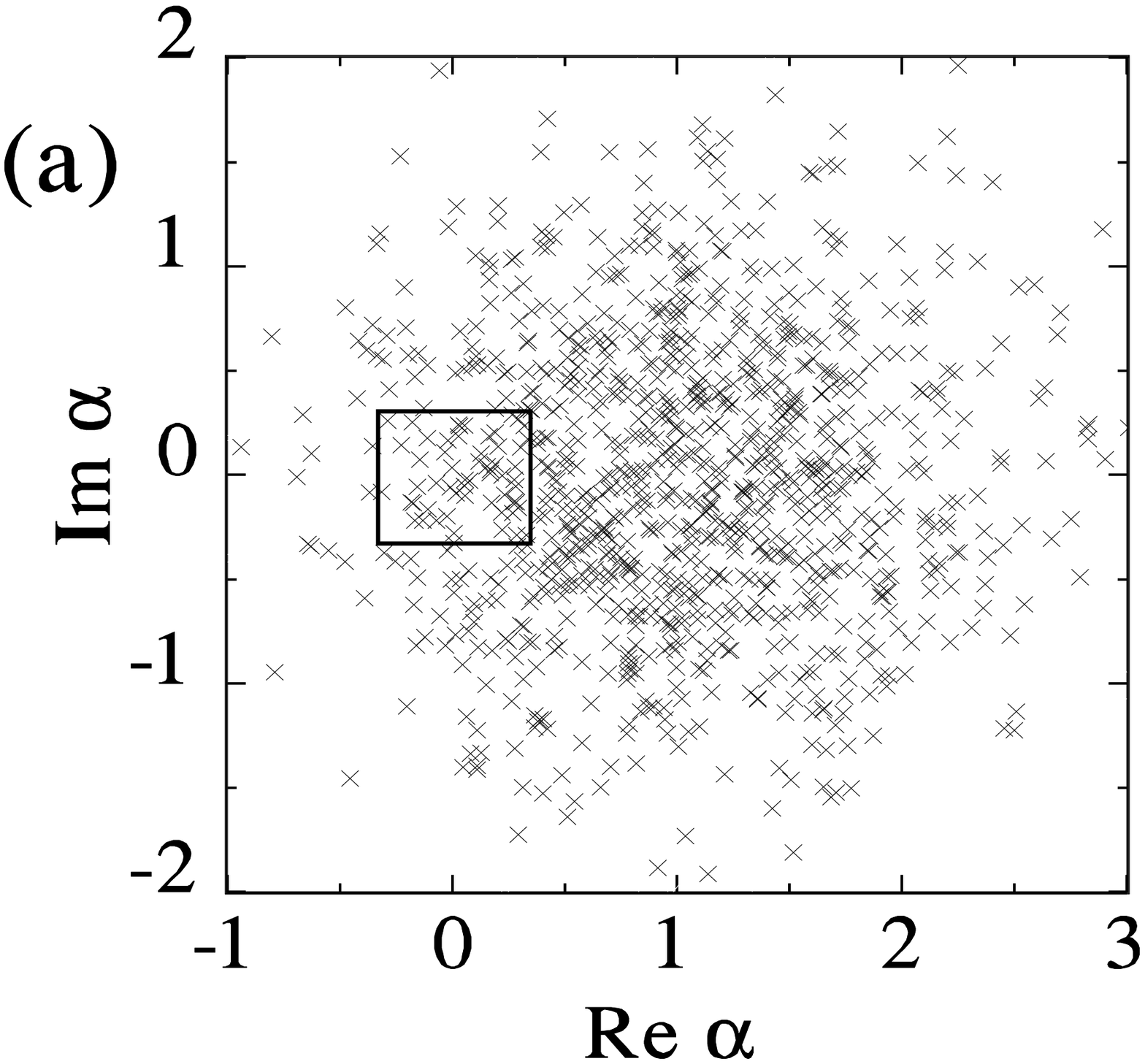,width=0.65\linewidth}}
\vspace*{2mm}
\centerline{\epsfig{figure=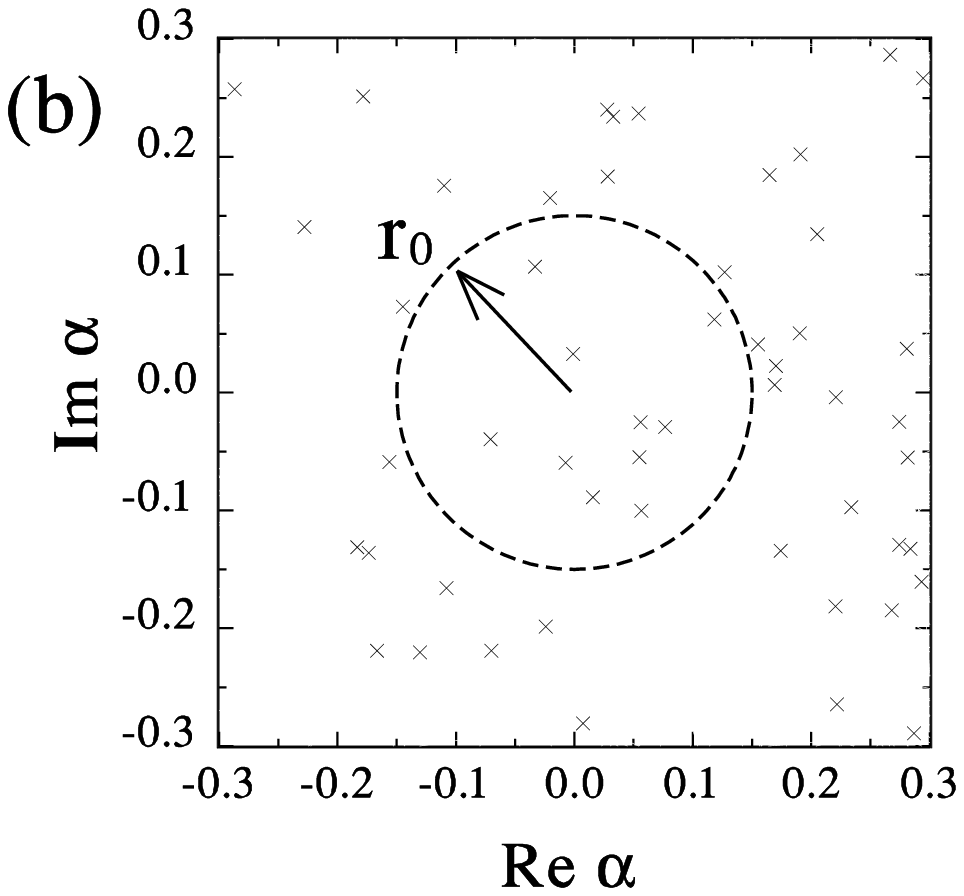,width=0.65\linewidth}}
\vspace*{2mm}
\caption{Output of simulated double-homodyne detection
of a coherent state $|\alpha\rangle$, $\alpha=1$ (a), and 
enlarged detail of the output around the origin (b).
}
\label{fig2}
\end{figure}


\subsection{Kernel regularization and sampling algorithm}

Since the main part of the statistical error arises from data close to the
origin, it is natural to modify the procedure by {\em omitting the data}
falling inside a small circle \mbox{$r$ $\!<$ $\!r_0$} (see Fig.~\ref{fig2}). 
Of course, such a deliberate data filtering introduces into the measurement
a state-dependent systematic error. Nevertheless, the statistical 
error is reduced and the total error may be acceptable. 
Replacing $K_k(r)$ by the regularized function $K^\prime_k(r)$ 
according to
\begin{equation}
 K^{\prime}_k(r)=\theta(r-r_0)K_k(r)
 \label{regulkernel}
\end{equation}
[$\theta(x)$, Heaviside step function], the systematic error of the
$k$th moment can be given by 
\begin{eqnarray}
 \sigma^{(\rm sys)}_k&=&\int_0^{2\pi} d\varphi \, e^{ik\varphi} 
 \int_0^{r_0} rdr\, Q(r,\varphi)K_k(r) 
 \nonumber \\
 &=&\int _0^{r_0} rdr\, Q_k(r)K_k(r) .
 \label{SSYS}
\end{eqnarray}
A measure of the total error is then the
sum of the statistical and systematic errors, 
\begin{eqnarray}
 {\rm Re}\, \sigma^{(\rm tot)}_k &=&
\left| {\rm Re}\, \sigma^{(\rm sys)}_k \right| +
 \left[D\!\left({\rm Re}\ \Psi_k^{(\rm est)}\right)\right]^{1/2},
\label{TOTERROR}
\end{eqnarray}
and ${\rm Im}\, \sigma^{(\rm tot)}_k$ accordingly.

\begin{figure}[!t!]
\centerline{\epsfig{figure=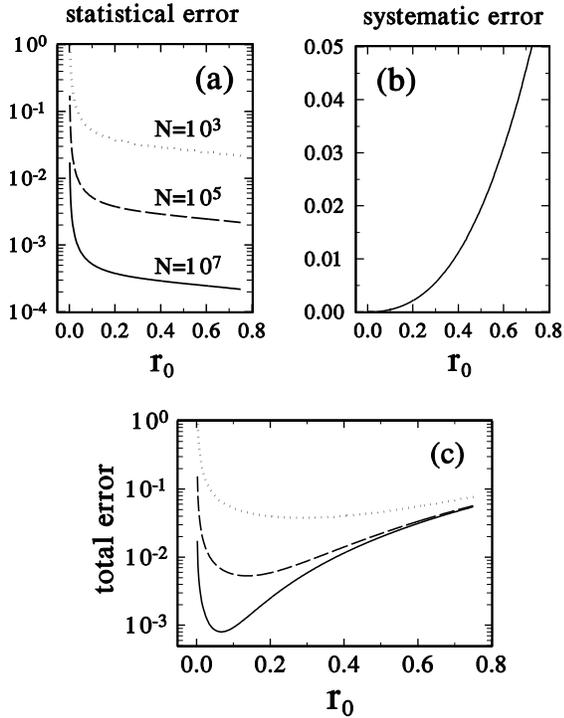,width=0.85\linewidth}}
\vspace*{2mm}
\caption{Statistical (a), systematic (b), and total (c) errors 
of the real part of the sampled exponential phase moment ${\rm Re}\Psi_2$ 
of a coherent state $|\alpha\rangle$, $\alpha$ $\!=$ $\!1$
for different numbers of recorded events $N$.
}
\label{fig2y}
\end{figure}

\begin{figure}[!t!]
\centerline{\epsfig{figure=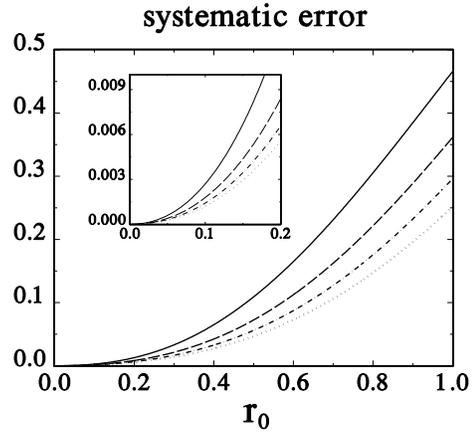,width=0.7\linewidth}}
\vspace*{2mm}
\caption{Upper bound of $|\sigma^{(\rm sys)}_k|$ estimated
from the inequality (\protect\ref{SYSSE})
for $k$ $\!=$ $\!1$ (solid line), $k$ $\!=$ $\!2$ (dashed line),
\mbox{$k$ $\!=$ $\!3$} (dot-dashed line), and $k$ $\!=$ $\!4$ (dotted line).}
\label{fig3}
\end{figure}

   From the example in Fig.~\ref{fig2y} it is seen that the
statistical error decreases with increasing radius $r_0$ 
[Fig.~\ref{fig2y}(a)], whereas the systematic error increases 
with the radius [Fig.~\ref{fig2y}(b)].
The total error has thus a minimum at a certain radius 
[Fig.~\ref{fig2y}(c)], which can be regarded as the optimal 
radius for regularization. Unfortunately, the determination
of the systematic error requires knowledge of the
state. Nevertheless, an upper bound of the systematic error
can be estimated, without any {\em a priori} knowledge of 
the measured state. Assuming $r$ $\!<$ $\!1$, we may write
\begin{eqnarray}
 \label{QKMAX}
\lefteqn{
 \left|Q_{k}(r)\right| 
 = 2\sum_{n=0}^{\infty}\frac{r^{2n+k}e^{-r^2}}{\sqrt{n!(n+k)!}}
 \left|\rho_{n+k,n}\right| 
} 
\nonumber\\&&\hspace{2ex}
 \,\leq 2\,\frac{r^k e^{-r^2}}{\sqrt{k!}}
 \sum_{n=0}^{\infty}\sqrt{\rho_{n,n}\,\rho_{n+k,n+k}}
\nonumber\\&&\hspace{2ex}
 \,\leq 2\,\frac{r^k e^{-r^2}}{\sqrt{k!}}
 \sum_{n=0}^{\infty}\frac{\rho_{n,n}+\rho_{n+k,n+k}}{2}
 \leq
 2\,\frac{r^k e^{-r^2}}{\sqrt{k!}}\,,
\end{eqnarray}
where we have used the inequality $|\rho_{m,n}|^2$ $\!\leq$ 
$\!\rho_{mm}\rho_{nn}$ implied by positive definiteness of $\rho$.
Hence, an upper bound of the systematic error can be estimated. 
Using Eq.~(\ref{SSYS}), we find that
\begin{equation}
 |\sigma^{(\rm sys)}_k| 
 \le \frac{2}{\sqrt{k!}}\int_{0}^{r_0} dr\, 
 r^{k+1} e^{-r^2} K_k(r).
 \label{SYSSE}
\end{equation}
A typical state for which $|\sigma^{(\rm sys)}_k|$ is
of the order of magnitude of upper-bound value 
is $|\psi_k\rangle$ $\!=$ $\!(|0\rangle$ $\!+$ $\!|k\rangle)/\sqrt{2}$. 
For this state, $Q_k(r)$ $\!=$ $\!r^k \exp(-r^2)/\sqrt{k!}$,
which yields one half of the upper bound value. 
Taking into account that $K_k(r)$ $\!\propto$ $\!1/r^k$ for 
$r$ $\!\ll$ $\!1$, we find from the inequality (\ref{SYSSE}) that 
the upper bound of $|\sigma^{(\rm sys)}_k|$ increases quadratically 
with $r_0$. The dependence on $r_0$ of the upper bound 
of the $|\sigma^{(\rm sys)}_k|$ is shown in Fig.~\ref{fig3}. Notice 
that the systematic error is smaller for higher $k$. 

The state-independent upper bound of the  systematic error 
and the estimated statistical error can now be used to
determine the upper bound of the total error. Its minimum
then determines an appropriate regularization radius $r_0$.
A possible algorithm for optimized data processing is the
following one. In the zeroth step, sampling of the desired
exponential phase moments from all $N$ measurement events
is performed. Since also data with very small $r$
may contribute to the result, the statistical error can be very large.
In contrast to standard sampling technique, where there is no need
for data storage, here the data within a certain small circle 
are stored. The radius of the circle should be slightly larger 
than the expected regularization radius. 
The regularized kernel function (\ref{regulkernel}) is now used, with 
$r_0$ being increased step by step, so that in the $n$th step $n$ 
events closest to the origin are covered by $r_0$.
In each step statistical and systematic errors are estimated.
The value of $r_0$ for which the total error is minimized is used for
calculation of the final result.

Let us mention that the detrimental effect of divergent kernels 
$K_k(r)$ at $r$ $\!=$ $\!0$ (in connection with nonzero $Q$-function)
resembles the experiment in \cite{Noh}, where the statistics of 
sine and cosine phases
are obtained from low-efficiency double homodyning.
In the experiment, data giving rise to divergences are disregarded, 
which is criticized in \cite{Hradil} from the argument
that the disregarded data represent an extra noise in the
statistics. In our case, we disregard data leading to high 
statistical error and include the resulting systematic error 
into the sampling scheme. 

\subsection{Total error and number of measurements}

Let us assume that a particular phase moment $\Psi_k$ is desired
to be determined with a prescribed total precision $\sigma^{(\rm tot)}_k$. 
What is the necessary number of measurement events $N$? 
If there were no need for regularization and the
precision were limited only by (finite) statistical fluctuation, 
then \mbox{$N$ $\!\propto$ $\!(\sigma^{(\rm tot)}_k)^{-2}$}. 
When the vacuum contributes to the state to be measured and
a regularization radius $r_0$ is introduced, then the total error reads
\begin{eqnarray}
\label{STOTOPTa}
 \sigma^{(\rm tot)}_1&=& A_1 (-\ln r_0)^{1/2}N^{-1/2}+B_1 r_0^2, 
\\
 \sigma^{(\rm tot)}_k&=& A_k r_0^{1-k} N^{-1/2}+B_k r_0^2, 
\qquad k\ge 2,
\label{STOTOPT}
\end{eqnarray}
where $A_k$ and $B_k$ are constants. 
The optimal regularization radius $r^{\rm (opt)}_0$, which 
minimizes the total error (\ref{STOTOPT}) depends on $N$ as 
\begin{equation}
r_0^{\rm (opt)} \propto N^{-1/[2(1+k)]}\,.
\label{STOTOPTb}
\end{equation}
{F}rom this expression and Eq.~(\ref{STOTOPT}) we find that
\begin{eqnarray}
 \sigma^{\rm (tot)} \propto N^{-1/(1+k)},  
 \label{SIGMAN}
\end{eqnarray}
i.e.,
\begin{eqnarray}
 N \propto \Big(\sigma^{\rm (tot)}_k\Big)^{-(1+k)}
\end{eqnarray}
($k$ $\!\ge$ $\!2$).
The case $k$ $\!=$ $\!1$ needs separate consideration, because of
the logarithm, which does not provide us with a simple analytical 
expression. Obviously, $N$ increases faster than 
($\sigma^{\rm (tot)}_1)^{-2}$ with decreasing error. 
Thus, we can see that in the limit of small total error ordinary homodyning 
(which does not require regularization) is better suitable for sampling 
exponential phase moments than the double homodyning, 
because it requires a smaller amount of data to achieve the same precision.

\begin{figure}
\epsfig{figure=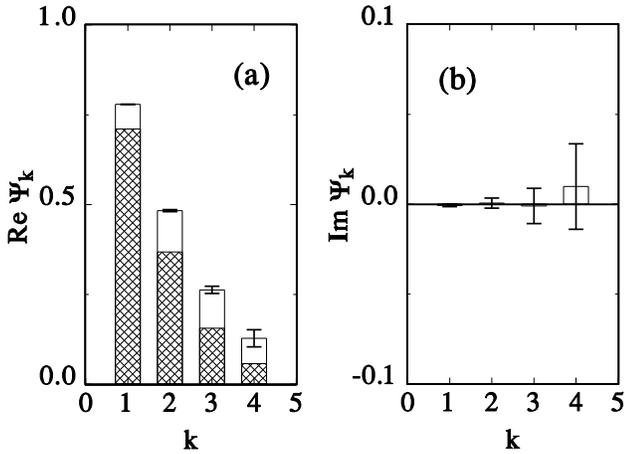,width=0.95\linewidth}
\vspace*{2.5mm}
\caption{Sampled exponential phase moments of a
coherent state $|\alpha\rangle$, $\alpha$ $\!=$ $\!1$; (a) real part of
$\Psi_k^{\rm (est)}$, (b) imaginary part of $\Psi_k^{\rm (est)}$. 
The error bars indicate the estimated statistical error. 
In the computer simulation, $N$ $\!=$ $\!10^6$ events are recorded 
and perfect detection is assumed.
The dashed regions correspond to the phase moments of the radially 
integrated $Q$-function.
}
\label{fig4}
\end{figure}


\subsection{Computer simulation}

To demonstrate the feasibility of the method, we have performed 
Monte Carlo simulations of double-homodyne detection of the 
$Q$-function for sampling the exponential phase moments of a 
coherent state.
Results are shown in Fig.~\ref{fig4} and Table 
\ref{tab1}. From Fig.~\ref{fig4} and Table \ref{tab1} it is seen that
the sampled exponential phase moments are in good agreement with the 
exact ones. Note the strong increase of the error with the index $k$ 
of the moment (for a detailed
error analysis, see Fig.~\ref{fig2y}).
Further, a comparison between the dashed and undashed bars in Fig.~\ref{fig4}
clearly shows the difference between the concept of 
canonical phase and the phase concept based on the 
radially integrated $Q$-function.

In order to compare double homodyning with ordinary homodyning, we 
have also simulated homodyne detection of the quadrature-component
statistics for sampling the exponential phase moments of the same
coherent state as in the simulated double-homodyne experiment, using the 
method in \cite{ODW98,DOW98,DBMOSW98}. The results are presented in
Table~\ref{tab2}.
Comparing Tables \ref{tab1} and \ref{tab2}, we see that
(for equal total numbers of events) the error in ordinary
homodyning is indeed smaller than in double homodyning. The
difference between the errors observed in the two schemes
increases with increasing index $k$ of the moment. Note
that for $k$ $\!=$ $\!4$ the error in the double-homodyne scheme 
is {\em ten times larger\/} than in the ordinary homodyne measurement.

\begin{table}[!t!]
\begin{tabular}{ccccc}
 $k$ & $\Psi_k$ & ${\rm Re} \, \Psi_k^{(\rm est)}$ &
${\rm Im}\,\Psi_k^{(\rm est)}$ &  $r_0$ \\
\hline
1 & 0.7732 & $0.7790\pm 0.0006$ & $-0.0008 \pm 0.0007~~$ & 0.007 \\
2 & 0.4805 & $0.483\pm 0.003$ & $0.001 \pm 0.003$ & 0.061 \\
3 & 0.2559 & $0.26\pm 0.01$ & $0.00 \pm 0.01$  & 0.160\\
4 & 0.1209 & $0.13\pm 0.02$ & $0.01 \pm 0.02$ & 0.277 \\
\end{tabular}
\vspace*{1.5mm}
\caption{Comparison of the sampled exponential phase moments
shown in Fig.~\protect\ref{fig4} with the exact ones. The displayed optimized
regularization radii $r_0$ refer to Re$\ \Psi_k$; values corresponding to 
Im$\ \Psi_k$ are similar in magnitude.
}
\label{tab1}
\end{table}

\begin{table}[!t!]
\begin{tabular}{cccc}
 $k$ & $\Psi_k$ & ${\rm Re} \, \Psi_k^{(\rm est)}$ &
${\rm Im}\,\Psi_k^{(\rm est)}$\\
\hline
1 & 0.7732 & $0.7736 \pm 0.0004$ & $0.0002 \pm 0.0006~~$ \\
2 & 0.4805 & $0.4795 \pm 0.0009$ & $-0.0003 \pm 0.001$ \\
3 & 0.2559 & $0.2573 \pm 0.0017$ & $0.0005 \pm 0.0017$ \\
4 & 0.1209 & $0.1200 \pm 0.0021$ & $0.0002 \pm 0.0021$ \\
\end{tabular}
\vspace*{1.5mm}
\caption{Comparison of the exponential phase moments
of a coherent state $|\alpha\rangle$, $\alpha$ $\!=$ $\!1$,  
sampled in homodyne detection with the exact ones. 
In the computer simulation, the $2\pi$-phase interval
of the quadrature components is divided into 120 equidistant 
values, and altogether $N$ $\!=$ $\!10^6$ events are recorded.
}
\label{tab2}
\end{table}


\subsection{Imperfect photodetection}

In a real experiment, the overall detection efficiency  $\eta$ would be always 
smaller than $100\%$, but it can be very high, e.g., $\eta$ $\!=$ $\!99\%$. 
Nonperfect detection introduces additional noise into the
sampling scheme and gives rise to an additional systematic error, 
which cannot be diminished by increasing the number of measurements. 
The effect of nonperfect detection is that the
exponential phase moments of a ``smoothed'' quantum state 
are sampled rather than those of the true one. 
Since the additional noise is Gaussian, the phase-space function that
is actually recorded is not the $Q$-function but the function 
$W_{1-2\eta^{-1}}(q,p)$. The sampling of $\Psi_k$ from 
this function by means of the kernel function
$K_k(r;-1)$ is equivalent to sampling of $\Psi_k$ from the
$Q$-function by means of the kernel function $K_k[r;-1+2(\eta^{-1}-1)]$. 
For a given quantum state, the systematic error can thus be given by
\begin{equation}
\Delta_{\eta}\Psi_k=
\int_0^{\infty} \!\! rdr\, Q_k(r)\left[K_k(r;-1)-K_k(r;-3+2\eta^{-1})\right].
\label{ETASYST}
\end{equation}
Its result is the underestimation of the magnitude of the moment.
Since the difference of the kernel functions is essentially nonzero 
only around the origin $r$ $\!=$ $\!0$, the systematic error 
$\Delta_\eta\Psi_k$
will be highest for states close to vacuum. 
After a proper regularization, one can use Eq.~(\ref{ETASYST})
to get a reasonable estimation of the error by substituting
the measured statistics for the unknown $Q$-function. 


\section{Other measurements of the phase-space functions}
\label{Other}

The phase moments $\Psi_k$ can be also obtained
from quasidistributions reconstructed
in unbalanced homodyning \cite{Wallentowicz96}, or cavity
measurements \cite{Lutterbach,Bodendorf}.
However, these methods 
do not yield $W_s(q,p)$ as statistics of events $(q,p)$,
but the functions $W_s(q,p)$ are determined pointwise.
The restriction in practice to a selected finite 
number of points necessarily results in a systematic error, because the
integration over the phase space is replaced by summation over a finite 
number of points selected by the experimentalist. 
Having determined $W_s(q,p)$, the exponential phase moments $\Psi_k$
can then be reconstructed from $W_s(q,p)$ on the basis of Eq.~(\ref{PSI}).
Since the kernel function $K_k(r;s)$ is well behaved for $s$ $\!>$ $\!-1$, 
no problems with divergences arise here.

In unbalanced homodyning displaced Fock-state distributions 
$p(n,\alpha)$ are measured \cite{Wallentowicz96,Quasi-n}. 
It can be expected that the cumbersome way of reconstructing $\Psi_k$
from $p(n,\alpha)$ via $W_s(q,p)$ may be avoided and the reconstruction 
can be performed directly from the measured data. This
could be done in a similar way as in the reconstruction of the  
density matrix in the Fock basis \cite{Opatrny97}.
A similar approach can be used for different physical systems: 
statistics of the displaced Fock states of vibrating trap ions
has been obtained in state-reconstruction experiments
\cite{Leibfried}, and schemes based on displaced Fock statistics
of the cavity fields have been suggested \cite{Lutterbach,Bodendorf}.
In particular, the scheme of \cite{Lutterbach}
directly yields the Wigner function, from which the exponential
phase moments can be obtained in a very straightforward
way. 
Even though many interesting problems are related to these 
schemes, we will not deal with them in this paper in 
any more detail.


\section{Phase distribution}
\label{Sec-distr}

In order to answer the question of the possibility of direct sampling 
of the phase distribution itself, we have first to answer the question of 
the existence of kernels \mbox{$F(r,\varphi-\psi;s)$} such that
\begin{equation}
  P(\varphi)=\int_0^{2\pi}d\psi\int_0^\infty 
  rdr \, W(r,\psi;s) F(r,\varphi-\psi;s).
  \label{PFW}
\end{equation}
Obviously, $F(r,\psi-\varphi;s)$ is the ($-s$)-parametrized phase-space
function
of the phase state $|\varphi\rangle$ in Eq.~(\ref{2}). In \cite{Eiselt91}
it is shown that this function can be given by
\begin{equation}
F(r,\varphi;s)=\sum_{m,n}B_{m,n}(r,s)e^{i(m-n)(\varphi)},
\label{FSERIES}
\end{equation}
where
\begin{eqnarray}
\lefteqn{
B_{mn}(r,s)=\sqrt{\frac{n!}{m!}} \, r^{m-n} \left(\frac{s-1}{2}\right)^n
\left(\frac{2}{1+s}\right)^{m+1}
}
\nonumber \\[.5ex]&&\hspace{12ex} \times \,
L_{n}^{m-n}\!\left(\frac{4r^2}{1-s^2}\right)\,
\exp\!\left(-\frac{2r^2}{1+s}\right)
\end{eqnarray}
for $m$ $\!\geq$ $\!n$, and $B_{nm}$ $\!=$ $\!B_{mn}$.
The series (\ref{FSERIES}) only converges for
$s$ $\!>$ $\!0$ (for the limiting case $s$ $\!=$ $\!0$, 
see also \cite{Herzog93}).

\begin{figure}
\centerline{\epsfig{figure=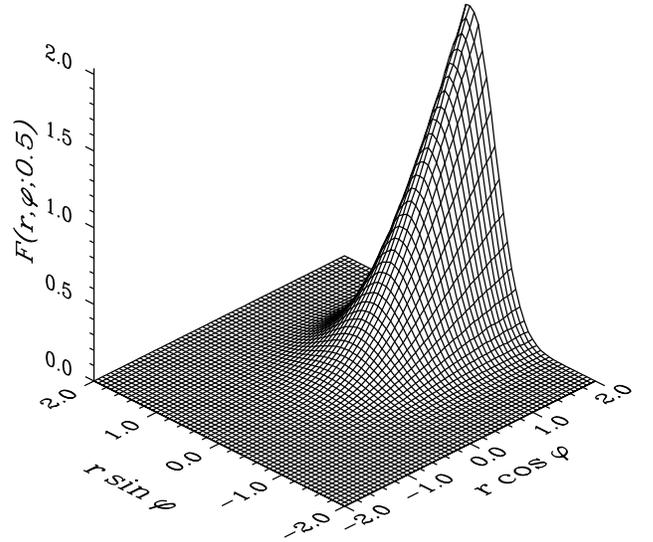,width=0.95\linewidth}}
\vspace*{2mm}
\caption{Plot of the function $F(r,\varphi;0.5)$.}
\label{figstat}
\end{figure}

Let us express $F(r,\varphi;s)$ for $s$ $\!>$ $\!0$ in terms of
$K(r;s)$. From Eq.~(\ref{defpsik}) it follows that
\begin{equation}
  \label{4}
  P(\varphi)=
  \frac{1}{2\pi}\sum_{k=-\infty}^\infty \Psi_k e^{-ik\varphi}.
\end{equation}
Combining Eqs.~(\ref{4}) and (\ref{PSI}) and recalling Eq.~(\ref{PFW}), 
we may write $F(r,\varphi-\psi;s)$ in the form
\begin{equation}
  F(r,\varphi;s)=\frac{1}{2\pi}
  \left[1+2\sum_{k=1}^{\infty}K_k(r;s)\cos(k\varphi)\right].
  \label{sumker}
\end{equation}
When $r$ $\!\rightarrow$ $\!\infty$ then $K_k(r;s)$ $\!\rightarrow$ $\!1$, 
and thus $F(r,\varphi,s)$ $\!\rightarrow$ $\!\delta(\varphi)$. 
Regrouping the terms in Eqs.~(\ref{KSUM}) and (\ref{CoefC})
\mbox{(for $s$ $\!>$ $\!0$)} and using a summation formula 
for Laguerre polynomials \cite{Prudnikov}, we can rewrite
$K_k(r;s)$ as
\begin{eqnarray}
\lefteqn{
  K_k(r;s)=r^k\left(\frac{2}{s\!+\!1}\right)^{k+1}e^{-2r^2/(1+s)}
}  
  \nonumber \\[.5ex]&&\hspace{1ex}\times\,
  \sum_{n=0}^\infty
  \frac{(-1)^n}{\sqrt{(n\!+\!1)\ldots(n\!+\!k)}}
  \left(\frac{1\!-\!s}{1\!+\!s}\right)^{n}
  L_n^k\!\left(\frac{4r^2}{1\!-\!s^2}\right),
  \label{KSER}
\end{eqnarray}
which is suitable for computing $F(r,\varphi;s)$. An example 
is displayed in Fig.~\ref{figstat}.

The fact that for a large class of states $W(r,\varphi;s)$ does not exist 
as a regular function for $s$ $\!>$ $\!0$ limits the applicability of 
Eq.~(\ref{PFW}). Nevertheless, there exist states for which $W(r,\varphi;s)$
for $s$ $\!>$ $\!0$ is a regular function which can be sampled using
unbalanced homodyning.
However, for $s$ $\!>$ $\!0$ the statistical error of the sampled 
$W(r,\varphi;s)$ increases with $r$ \cite{Banaszek2}, so that
application of Eq.~(\ref{PFW}) requires special regularization.


\section{Summary and Conclusion}
\label{Sec-conclusion}

The main results can be summarized as follows.
$(i)$ There exist well-behaved kernels for sampling the exponential 
moments of the canonical phase from $s$-parametrized phase-space functions
for $s$ $\!>$ $\!-1$. For $s$ $\!=$ $\!-1$ the kernels diverge in the
origin, and for $s$ $\!<$ $\!-1$ the kernels do not exist as regular 
functions. 
$(ii)$ Even though for $s$ $\!=$ $\!-1$ the kernels diverge, their 
integral with the $Q$-function is finite, so that they may be used 
for inferring the exponential phase moments from the exact $Q$-function. 
However, the kernel divergence may cause divergent errors of some moments 
for some states if fluctuating experimental data are used.
$(iii)$ 
Finite errors can be obtained if regularized kernels are used. Since
regularization introduces a systematic error, an optimization 
procedure should be used in order to minimize the sum of the 
statistical and systematic errors.
$(iv)$
The fact that the canonical phase moments can be sampled in 
double homodyning has an interesting interpretation. 
Each measurement event yields 
a unique phase value, but these values must be taken with different weights 
in dependence on the distance from the origin of the phase space.
This is in contrast to the ordinary (four-port) homodyning, where a single
measurement does not provide us with a phase value.
$(v)$ 
Even if optimally regularized kernels are used, 
the amount of data necessary for realizing a desired precision
is larger than in standard sampling. 
This is a disadvantage of the double-homodyne scheme in 
comparison with ordinary  homodyning.
$(vi)$ 
In double homodyning, correct results require
perfect detection, because there is no simple possibility of compensation 
for detection losses, which cause an additional systematic error. 
This is another disadvantage of double homodyning compared to
ordinary homodyning where a compensation of imperfect 
detection is possible for efficiencies down to $\eta$ $\!>$ $\!0.5$.
$(vii)$ 
Thus, in reply to the question posed in the Introduction, it does not 
seem that phase-space measurements based on double-homodyning are closer to 
canonical-phase measurement than quadrature-component measurements based 
on ordinary homodyning.
$(viii)$ In contrast to ordinary homodyning however, the sampling functions
in double homodyning are uniquely defined. This follows from the 
fact that they are actually phase-space representations of 
quantum mechanical operators.
$(ix)$ The exponential phase moments can also be 
inferred from the data recorded in other schemes
such as unbalanced homodyning, in which $s$-parametrized 
phase-space functions are reconstructed pointwise.
$(x)$ Kernels for sampling the distribution of the canonical phase exist 
as regular functions only for
$s$ $>$ 0. Even though for some states the corresponding phase space functions
exist and can be measured using unbalanced homodyning, 
the behavior of the statistical error would require a special
regularization of the 
scheme to be applicable.


\acknowledgments

We thank J. Pe\v{r}ina for stimulating discussions and
acknowledge discussions with J. Clausen.
J.F. acknow\-ledges support of the US-Israel Binational Science Foundation.
This work was supported by the Deutsche Forschungsgemeinschaft.

\appendix


\section{Sampling kernels and  the function $\Omega$}
\label{derivkernel}


Substituting Eq.~(\ref{MOMENTSA}) into Eq.~(\ref{PSIaa}) and comparing 
with Eq.~(\ref{PSI}), we can express $K_k(r;s)$  as
\begin{equation}
 K_k(r;s)=r^k\sum_{l=0}^{\infty}\left(\frac{1-s}{2}\right)^l
 C_l^{(k)} L_l^k\!\left(\frac{2r^2}{1-s}\right),
 \label{KSUM}
\end{equation}
where the coefficients
\begin{eqnarray}
 C_l^{(k)} = \sum_{n=0}^{l}\frac{(-1)^n}{\sqrt{(n\!+\!1)\dots
 (n\!+\!k)}} {l \choose n} ,
 \label{CoefC}
\end{eqnarray}
can be rewritten as
\begin{equation}
C_l^{(k)}=\frac{1}{\pi^{k/2}}
\int_{-\infty}^{\infty}dt_1\, e^{-t_1^2}\ldots 
\int_{-\infty}^{\infty}dt_k\, e^{-kt_k^2}z_k^l ,
\label{CINT}
\end{equation}
with
\begin{equation}
z_k=1-e^{-\rho_k^2},\qquad
\rho_k^2=\sum_{j=1}^{k}t_j^2.
\label{Z}
\end{equation}
Substituting this expression into Eq.~(\ref{KSUM}) 
and using the summation rule
\begin{eqnarray}
\lefteqn{
 \sum_{l=0}^{\infty}
 \left(\frac{1-s}{2}z_k\right)^l L_l^k\!\left(\frac{2r^2}{1-s}\right)
} 
 \nonumber \\[.5ex]&&\hspace{2ex}
 = \left(1-z_k\frac{1-s}{2}\right)^{-k-1}
 \exp\!\left[\frac{z_k r^2}{z_k(1-s)/2-1}\right],
 \label{SUM}
\end{eqnarray}
we arrive at
\begin{eqnarray}
\lefteqn{
K_k(r;s)=\frac{r^k 2^{k+1}}{\pi^{k/2}}
\int_{-\infty}^{\infty}dt_1\, \bigg\{ e^{-t_1^2}\ldots 
\int_{-\infty}^{\infty}dt_k\, e^{-kt_k^2}
}
\nonumber \\[.5ex]&&\hspace{1ex}
\times\left[1\!+\!s\!+\!(1\!-\!s)e^{-\rho_k^2}\right]^{-\!k\!-\!1}
\exp\!\bigg[-\frac{2(1\!-\!e^{-\rho_k^2})r^2}{1\!+\!s\!
+\!(1\!-\!s)e^{-\rho_k^2}}\bigg]\bigg\}
\nonumber\\&&
\label{K}
\end{eqnarray}
Note that the series in Eq.~(\ref{SUM}) is only convergent for 
\mbox{$|z_k(1-s)/2|<1$}. We have $z_k$ $\!\leq$ $\!1$, thus
\begin{equation}
|(1-s)/2| < 1 \quad \Rightarrow \quad s > -1
\end{equation}
must hold so that the $Q$-function ($s$ $\!=$ $\!-1)$ represents limiting
case for sampling the phase moments.

The multiple integration in Eq.~(\ref{K}) 
can be conveniently performed in hyperspherical
coordinates. For this purpose we introduce the function 
$\Omega^{(k)}(\rho^2)$ \cite{DOW98},
\begin{eqnarray}
\lefteqn{
 \Omega^{(k)}(\rho^2)=\int_0^\pi d\varphi_1\,\bigg[\sin^{k-2}\varphi_1\ldots
 \int_0^\pi d\varphi_i\,\sin^{k-i-1}\varphi_i
} 
 \nonumber \\&&\hspace{12ex}\times\,
 \ldots
 \int_0^{2\pi} d\varphi_{k-1}\,\exp\!\bigg(-\sum_{l=1}^k l t_l^2\bigg)\bigg],
 \label{B1}
\end{eqnarray}
where 
\begin{eqnarray}
t_{i}=\rho \cos \varphi_{i} \prod_{j=1}^{i-1}\sin \varphi_{j} 
\quad {\rm if}
\quad i<k ,
\end{eqnarray}
and
\begin{eqnarray}
t_{k}=\rho \prod_{j=1}^{k-1}
\sin \varphi_{j} ,
\end{eqnarray}
with $\rho$ $=$ $\rho_k$.
The exponent can be expressed in hyperspherical coordinates as
\begin{eqnarray}
\lefteqn{
\sum_{l=1}^k lt_l^k = \rho^2 
}
\nonumber\\&&\hspace{1ex}
+ \,\rho^2\sin^2\varphi_1(1\!+\!\sin^2\varphi_2
(1\!+\!\ldots(1\!+\!\sin^2\varphi_{k-1}))).
\label{BEXP}
\end{eqnarray}
Inserting this expression into Eq.~(\ref{B1}) and 
expanding the exponential function
into a Taylor series, we arrive at Eq.~(\ref{OMEGATAYLOR})
together with Eq.~(\ref{5}).
A recurrence formula for the coefficients $A_{n}^{(k)}$ in
Eq.~(\ref{5}) can be readily obtained:
\begin{equation}
A_{n}^{(k)}=B_{2n+k-2}\sum_{l=0}^n {n \choose l} A_l^{(k-1)},
\quad k\ge 3
\label{AREC}
\end{equation}
where 
\begin{equation}
B_{j}=\int_0^\pi d\varphi\,\sin^{j}\varphi
=\sqrt{\pi}\frac{\Gamma(\frac{j+1}{2})}
{\Gamma(\frac{j+2}{2})},\quad (j\ge 0).
\end{equation}
Starting from $A_n^{(2)}$ $\!=$ $\!2B_{2n}$, the formulas (\ref{AREC}) and 
(\ref{OMEGATAYLOR}) allow for fast and accurate numerical 
determination of the functions $\Omega^{(k)}(\rho^2)$
even for high $k$.


\section{Asymptotics of $\Omega^{({\rm k})}(\rho^2)$ and divergence of
kernels $K_{\rm k}(r)$}

In order to analyze the divergence of the kernels $K_k(r)$,
we must first know the asymptotic behavior
of the functions $\Omega^{(k)}(\rho^2)$ for large $\rho$.
We start from  the integral representation (\ref{B1})
and write the exponent (\ref{BEXP}) as
\begin{equation}
\sum_{l=1}^k lt_l^k=\rho^2+\rho^2\sin^2\varphi_1
\Phi(\varphi_2,\ldots,\varphi_{k-1}),
\label{EXPONENT}
\end{equation}
with
\begin{equation}
\Phi(\varphi_2,\ldots,\varphi_{k-1})=
1+\sin^2\varphi_2(1+\sin^2\varphi_3(1+\ldots)).
\end{equation}
Note that $\Phi$ $\!\geq$ $\!1$.
We insert Eq.~(\ref{EXPONENT}) into Eq.~(\ref{B1}) and integrate 
over $\varphi_1$.
Assuming $k\ge 2$, 
the relevant integral is
\begin{eqnarray}
I&=&\int_0^\pi d\varphi_1\,
\sin^{k-2}\varphi_1 \,e^{-\rho^2\Phi\sin^2\varphi_1 }
\nonumber \\
&=&2\int_0^{\frac{\pi}{2}}d\varphi_1\, 
\sin^{k-2}\varphi_1 \,e^{-\rho^2\Phi\sin^2\varphi_1 }.
\end{eqnarray}
Assuming $\rho^2$ $\!\gg$ $\!1$, we may write $\sin\varphi_1$ $\!\approx$
$\!\varphi_1$, because the integrand is essentially nonzero only for
$\varphi_1$ $\!\ll$ $\!1$. From the same argument, we can extend the
integration from $(0,\pi/2)$ to $(0,\infty)$,
\begin{equation}
I\approx 2\int_0^\infty d\varphi_1 \, 
\varphi_1^{k-2} e^{-\rho^2\Phi\varphi_1^2}
=\Gamma\left(\frac{k-1}{2}\right)(\rho^2\Phi)^{\frac{1-k}{2}}.
\end{equation}
To finish the calculation of $\Omega$,
one has to integrate $\Phi^{(1-k)/2}$ over the remaining angles
$\varphi_2,\ldots,\varphi_{k-1}$ (with appropriate measure).
We eventually find the asymptotic behavior 
\begin{equation}
\Omega^{(k)}(\rho^2)\sim C_k \rho^{1-k} e^{-\rho^2}.
\label{OMEGAAS}
\end{equation}
The factor $\exp(-\rho^2)$ comes from the first $\rho^2$ in 
Eq.~(\ref{EXPONENT}). 
Taking into account that $\Omega^{(1)}(\rho^2)$ $\!=$
$\!2\exp(-\rho ^2)$,
we see that the asymptotic behavior (\ref{OMEGAAS})
holds for all $k\ge 1$.

To investigate the divergence of
$K_k(r)$ at $r\rightarrow 0$, we
make use of the integral representation (\ref{KERNEL}),
which is rewritten here as
\begin{eqnarray}
\lefteqn{
 K_k(r)=\frac{r^k}{\pi^{k/2}}
 \int_{0}^{\infty} d\rho \,\rho^{k-1} \Omega^{(k)}(\rho^2)
} 
 \nonumber \\[.5ex]&&\hspace{12ex} \times\, 
 e^{(k+1)\rho^2}
 \exp\!\left[-(e^{\rho^2}-1)r^2\right].
 \end{eqnarray}
For small $r$, the dominant contribution to this
integral comes from large $\rho$.
We can replace $\Omega$ by  the asymptotic formula (\ref{OMEGAAS})
and absorb the prefactors into $C_k$,
\begin{equation}
K_k(r)\approx C_k r^k\int_{0}^{\infty}  d \rho\, e^{k \rho^2}
\exp\!\left[-(e^{\rho^2}-1)r^2\right].
\label{KASINT}
\end{equation}
Change of the variable according to
\begin{equation}
t=(e^{\rho^2}-1)r^2
\end{equation}
yields
\begin{equation}
K_k(r)\approx\frac{1}{2r^k}\int_0^{\infty} d t
\left(\ln\frac{t+r^2}{r^2}\right)^{-1/2}  (t+r^2)^{k-1}
e^{-t}.
\label{B9}
\end{equation}
The integration region of (\ref{B9}) can be divided in
two parts, $t$ $\!<R^2$ and $t$ $\!>R^2$, with $R$ $\!\approx$ $\!r$.
For small $r$, the dominant contribution stems from the latter part
where the approximation
 \begin{equation}
\ln\frac{t+r^2}{r^2}\approx \ln\frac{1}{r^2}
 \end{equation}
can be used, and we find that
\begin{equation}
K_k(r)\approx\frac{(-2\ln r)^{-1/2}}{2r^k}\int_0^{\infty} d t
(t+r^2)^{k-1}e^{-t}.
\end{equation}
The integral is finite for $r$ $\!\rightarrow$ $\!0$, and thus
\begin{equation}
K_k(r)\sim \frac{r^{-k}}{(-\ln r)^{1/2}}.
\label{KFULLDIVER}
\end{equation}
The logarithm singularity in the denominator is
very weak in comparison to the polynomial one in the numerator. Only 
the polynomial divergence is relevant in Eq.~(\ref{PSI}),
because it determines whether the integral is convergent or not.
Thus we need not consider the logarithmic part, so that 
Eq.~(\ref{KFULLDIVER}) simplifies to
\begin{equation}
K_k(r) \sim r^{-k}.
\label{KDIVER}
\end{equation}



\begin{references}
\bibitem{Quasi-n}
Alternatively, $p(n,\alpha)$ with a fixed $n$ can be treated as a generalized
phase space function based on scalar product of the signal state with a
displaced Fock state. 
Such a function is another representation of the signal
state.
\end{references}
\end{document}